\documentclass[final,12pt]{elsarticle}

\usepackage{amsmath,amssymb}
\usepackage{booktabs}
\usepackage{graphicx}
\usepackage{float}
\usepackage{placeins}
\usepackage{hyperref}
\usepackage{array}
\usepackage{enumitem}

\journal{Journal of Biomedical Informatics}

\begin{document}

\begin{frontmatter}

\title{Entropy-Dominated Temporal Vocal Dynamics as Digital Biomarkers for Depression Detection}

\author{Himadri Sekhar Samanta}
\address{Independent Researcher, Austin, Texas, USA}
\ead{himadri.tphysics@gmail.com}

\begin{abstract}
Automated depression detection often relies on static aggregation of conversational signals, potentially obscuring clinically meaningful behavioral dynamics. We investigated whether entropy-driven temporal biomarkers improve depression detection beyond standard pooled features using the DAIC-WOZ corpus. Using 142 labeled participants, we reconstructed utterance-level acoustic trajectories and compared pooled temporal baselines, trajectory dynamics, Shannon entropy biomarkers, recurrence quantification, sample entropy, fractal complexity, and coupling biomarkers under leakage-aware validation. Static pooling achieved an AUC of 0.593, trajectory dynamics improved performance to 0.637, and entropy biomarkers produced the strongest statistically significant improvement over pooled baselines (AUC 0.646; nested cross-validated AUC 0.615; permutation $p=0.017$). Entropy biomarkers outperformed recurrence, coupling, sample entropy, and fractal-based features, with several biomarkers stable across folds. These findings suggest depression-related signal may lie less in average acoustic levels than in entropy of conversational dynamics, supporting temporally informed digital phenotypes for mental-health assessment.
\end{abstract}

\begin{keyword}
Depression detection \sep Digital phenotyping \sep Entropy biomarkers \sep Conversational speech \sep Temporal dynamics \sep Biomedical informatics
\end{keyword}

\end{frontmatter}

\section{Introduction}

Major depressive disorder remains a leading contributor to disability worldwide, motivating scalable and objective computational screening methods. Conversational behavior provides rich acoustic and behavioral signals relevant to affective dysregulation, psychomotor change, cognitive load, and digital phenotyping \cite{cummins2015review,low2020automated,insel2017digital,onnela2016harnessing}.

The Distress Analysis Interview Corpus--Wizard of Oz (DAIC-WOZ) is a widely used benchmark for automated depression analysis from semi-structured clinical interviews \cite{gratch2014daic,valstar2016avec}. Prior work has explored acoustic features, lexical representations, multimodal fusion, recurrent sequence models, and deep architectures for depression detection \cite{alhanai2018detecting,ma2016depaudionet,haque2018measuring}. However, many pipelines still summarize time-varying signals using static aggregation such as means and standard deviations over an interview. Such summaries are useful but may obscure clinically meaningful temporal organization.

Depression may alter not only average behavior, but also variability, uncertainty, persistence, and temporal coordination. These properties are naturally framed as dynamical biomarkers rather than static descriptors. Entropy, recurrence structure, fractal dynamics, and nonlinear complexity have been informative in physiological and behavioral systems \cite{shannon1948mathematical,marwan2007recurrence,goldberger2002fractal,higuchi1988approach,richman2000sample}. Yet their relative utility for conversational depression detection has not been systematically compared under leakage-aware evaluation.

Unlike prior work emphasizing architectural complexity, this study investigates which family of temporal behavioral descriptors carries robust depression signal under leakage-aware validation.

This work makes three primary contributions. First, it develops a temporal biomarker framework for depression detection from conversational trajectories. Second, it demonstrates that entropy-dominated vocal biomarkers outperform static aggregation and several alternative dynamical complexity families. Third, it combines leakage-aware validation with translational deployment framing for clinical decision support.

\section{Related Work}

\subsection{Speech and multimodal depression detection}

Automated depression detection from speech has been studied using prosodic, lexical, spectral, and multimodal signals \cite{cummins2015review,williamson2013vocal,low2020automated}. DAIC-WOZ and AVEC challenges enabled standardized evaluation of depression recognition systems \cite{gratch2014daic,valstar2016avec,ringeval2019avec}. Deep models such as recurrent neural networks and multimodal fusion models have also been explored \cite{alhanai2018detecting,ma2016depaudionet,yang2017depression}. However, small labeled clinical datasets make interpretability and overfitting control especially important \cite{hastie2009elements,pedregosa2011scikit}.

\subsection{Temporal dynamics and complexity biomarkers}

Complexity measures have long been used to characterize physiological systems where disease and aging may alter structured variability \cite{goldberger2002fractal,pincus1991approximate,richman2000sample}. Shannon entropy quantifies distributional uncertainty \cite{shannon1948mathematical}; recurrence quantification analysis characterizes repeated states \cite{marwan2007recurrence}; and Higuchi fractal dimension estimates irregular geometric complexity \cite{higuchi1988approach,kantz2004nonlinear}. This study compares these biomarker families under a common depression-detection framework.

\subsection{Biomedical informatics and deployment}

Clinical machine-learning systems require not only predictive performance, but also calibration, monitoring, interpretability, and safe human oversight \cite{beam2018big,rajkomar2019machine,topol2019high,sendak2020path}. Digital mental-health tools additionally require attention to privacy, drift, demographic bias, and inappropriate autonomous use \cite{insel2017digital,onnela2016harnessing,shatte2019machine,dwyer2018machine}. We therefore include translational deployment considerations as part of the study design.

\section{Materials}

\subsection{Dataset}

We used DAIC-WOZ, a corpus of semi-structured clinical interviews designed for analysis of psychological distress \cite{gratch2014daic}. The local dataset contained 189 participant folders. The labeled train and development subset used for analysis contained 142 participants, including 42 depression-positive and 100 depression-negative participants based on the PHQ-8 binary label.

\begin{table}[H]
\centering
\caption{Dataset configuration used in this study.}
\begin{tabular}{lc}
\toprule
Quantity & Value \\
\midrule
Total participant folders available & 189 \\
Labeled train+development participants & 142 \\
Depression-positive participants & 42 \\
Depression-negative participants & 100 \\
Fixed utterance length after reconstruction & 150 \\
Voice features per utterance & 150 \\
Voice tensor shape & $189 \times 150 \times 150$ \\
\bottomrule
\end{tabular}
\end{table}

\subsection{Data integrity and reconstruction}

Initial inspection revealed that a previously aggregated segment file produced duplicated participant trajectories. Therefore, all main analyses were rebuilt from raw participant files rather than from the corrupted aggregate. Participant transcripts were used to identify participant utterance boundaries. COVAREP acoustic streams were aligned to each utterance using timestamp-based indexing, and utterance-level means and standard deviations were computed for each acoustic channel. Participants were represented as fixed-length sequences of 150 utterances. Shorter sequences were zero-padded; longer sequences were truncated to the first 150 participant utterances. Missing and non-finite values were replaced by zero after reconstruction.

This reconstruction step was essential because the primary study question concerns temporal dynamics. Any duplicated or incorrectly aligned tensor would invalidate downstream dynamical analysis.

\section{Methods}

\subsection{Problem formulation}

For participant $i$, let
\begin{equation}
X_i = \{x_{i1}, x_{i2}, \ldots, x_{iT}\},
\end{equation}
where $T=150$ utterances and $x_{it}\in \mathbb{R}^{150}$ is the acoustic feature vector for utterance $t$. The target label is $y_i\in\{0,1\}$, where 1 denotes depression-positive status.

\subsection{Static temporal pooling baseline}

For each feature trajectory $x^{(j)}_{1:T}$, we computed mean, standard deviation, and maximum:
\begin{equation}
\mu_j = \frac{1}{T}\sum_{t=1}^{T}x_t^{(j)},
\end{equation}
\begin{equation}
\sigma_j = \sqrt{\frac{1}{T}\sum_{t=1}^{T}(x_t^{(j)}-\mu_j)^2},
\end{equation}
\begin{equation}
M_j = \max_t x_t^{(j)}.
\end{equation}
This produced 450 static pooled features.

\subsection{Trajectory dynamical biomarkers}

For each trajectory, we extracted linear slope, standard deviation of first differences, volatility, trend reversal count, Shannon entropy, and lag-1 autocorrelation. First differences were defined as:
\begin{equation}
\Delta x_t = x_{t+1}-x_t.
\end{equation}
Volatility was:
\begin{equation}
V = Var(\Delta x_t).
\end{equation}
Lag-1 autocorrelation was:
\begin{equation}
\rho_1 = corr(x_t,x_{t+1}).
\end{equation}

\subsection{Entropy biomarkers}

Shannon entropy was computed after binning each trajectory:
\begin{equation}
H(X) = -\sum_{b=1}^{B}p_b\log(p_b),
\end{equation}
where $p_b$ is the empirical probability of values falling in bin $b$. Entropy features quantify distributional uncertainty in acoustic trajectories.

\subsection{Recurrence quantification analysis}

For each normalized trajectory, we constructed a recurrence matrix:
\begin{equation}
R_{ij} = \mathbf{1}(|x_i-x_j|\leq \epsilon),
\end{equation}
with $\epsilon$ set to 10\% of the maximum pairwise distance. Recurrence rate was:
\begin{equation}
RR = \frac{1}{T^2}\sum_{i,j}R_{ij}.
\end{equation}
A determinism proxy estimated the fraction of recurrence points participating in short diagonal structures.

\subsection{Sample entropy and fractal dimension}

Sample entropy was computed with embedding dimension $m=2$ and tolerance $r=0.2\sigma$ \cite{richman2000sample}. Higuchi fractal dimension was computed with $k_{max}=10$ \cite{higuchi1988approach}. These measures evaluated whether classical nonlinear complexity descriptors outperformed Shannon entropy.

\subsection{Cross-modal coupling}

As a first-pass multimodal analysis, gaze trajectories were aligned to participant utterances. Gaze variability was computed from $x_0$ and $y_0$ gaze components, and speech--gaze coupling was defined as the correlation between word-count trajectories and gaze variability. We also evaluated an entropy-coupling feature based on correlation between utterance-level speech-change magnitude and gaze entropy.

\subsection{Models and validation}

The primary model was regularized logistic regression with balanced class weights. Random forest and gradient boosting benchmarks were also evaluated. Logistic regression was retained as the primary model because it provided competitive performance and direct interpretability.

We used stratified five-fold cross-validation for primary model comparison. Hyperparameter-tuned nested cross-validation was used for the entropy model. A permutation test with 1000 permutations evaluated whether the observed entropy-model AUC exceeded chance under label shuffling. Out-of-fold predictions were used to compute balanced accuracy, sensitivity, specificity, and the confusion matrix.

\section{Results}

\subsection{Main model comparison}

\begin{figure}[H]
\centering
\includegraphics[width=0.85\linewidth]{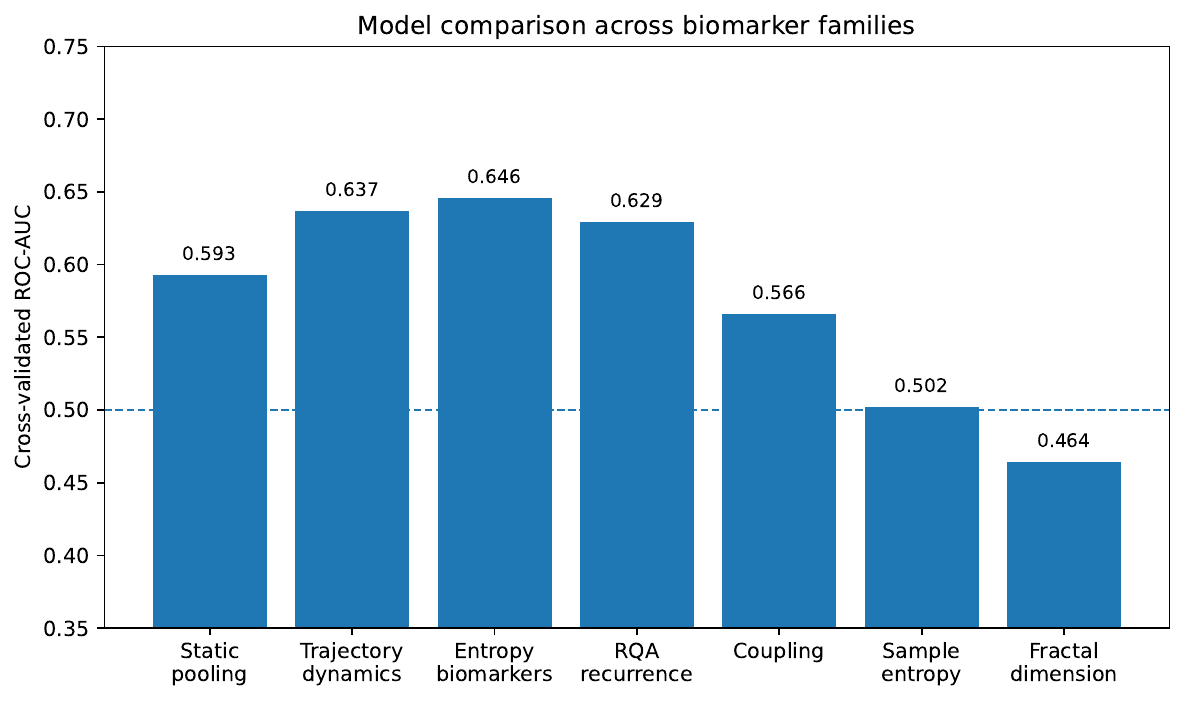}
\caption{Model comparison across biomarker families. Entropy-driven vocal biomarkers produced the strongest valid signal among evaluated feature families.}
\label{fig:auc}
\end{figure}

Table~\ref{tab:modelcomparison} and Figure~\ref{fig:auc} summarize the main results. Static pooling achieved mean AUC 0.593. Full trajectory dynamics improved performance to 0.637. Shannon entropy biomarkers achieved the strongest non-nested cross-validated AUC of 0.646. Nested cross-validation yielded a conservative entropy AUC of 0.615.

\begin{table}[H]
\centering
\caption{Model comparison across feature families. AUC values are mean cross-validated ROC-AUC unless otherwise stated.}
\label{tab:modelcomparison}
\begin{tabular}{lc}
\toprule
Feature/model family & AUC \\
\midrule
Static pooled baseline (mean/std/max) & 0.593 \\
Full trajectory dynamics & 0.637 \\
Shannon entropy biomarkers & 0.646 \\
Nested CV entropy model & 0.615 \\
RQA recurrence rate & 0.629 \\
RQA determinism proxy & 0.542 \\
Simple voice--gaze coupling & 0.566 \\
Entropy + coupling & 0.644 \\
Gradient boosting on entropy features & 0.605 \\
Sample entropy & 0.502 \\
Higuchi fractal dimension & 0.464 \\
\bottomrule
\end{tabular}
\end{table}

\subsection{Permutation significance}

The entropy model significantly exceeded a label-shuffled null distribution. Observed AUC was 0.646, compared with null mean 0.496 and null standard deviation 0.072 ($p=0.017$).

\begin{table}[H]
\centering
\caption{Permutation significance test for entropy biomarkers.}
\begin{tabular}{lc}
\toprule
Quantity & Value \\
\midrule
Observed AUC & 0.646 \\
Permutation null mean & 0.496 \\
Permutation null SD & 0.072 \\
Permutation p-value & 0.017 \\
Number of permutations & 1000 \\
\bottomrule
\end{tabular}
\end{table}

\subsection{Calibration and confusion matrix}

Out-of-fold classification using the entropy model produced balanced accuracy 0.612, sensitivity 0.405, and specificity 0.820.

\begin{table}[H]
\centering
\caption{Out-of-fold confusion matrix for the entropy model.}
\begin{tabular}{lcc}
\toprule
 & Predicted control & Predicted depressed \\
\midrule
True control & 82 & 18 \\
True depressed & 25 & 17 \\
\bottomrule
\end{tabular}
\end{table}

\subsection{Biomarker stability}

Feature-importance analysis showed that the strongest entropy markers were dominated by acoustic variability entropy rather than raw level features. To reduce post-hoc interpretation risk, we examined stability of top-ranked entropy biomarkers across five folds.

\begin{table}[H]
\centering
\caption{Most stable entropy biomarkers across cross-validation folds.}
\label{tab:stable}
\begin{tabular}{lc}
\toprule
Entropy biomarker & Top-10 fold frequency \\
\midrule
cov\_0.38\_std\_entropy & 5/5 \\
cov\_0.46\_std\_entropy & 4/5 \\
cov\_0.36\_mean\_entropy & 3/5 \\
cov\_0.135\_mean\_entropy & 3/5 \\
cov\_0.40\_std\_entropy & 3/5 \\
cov\_0.5\_std\_entropy & 3/5 \\
cov\_0.24\_std\_entropy & 3/5 \\
cov\_0.6\_std\_entropy & 3/5 \\
cov\_0.38\_mean\_entropy & 3/5 \\
cov\_0.2\_std\_entropy & 2/5 \\
\bottomrule
\end{tabular}
\end{table}

\begin{figure}[H]
\centering
\includegraphics[width=0.85\linewidth]{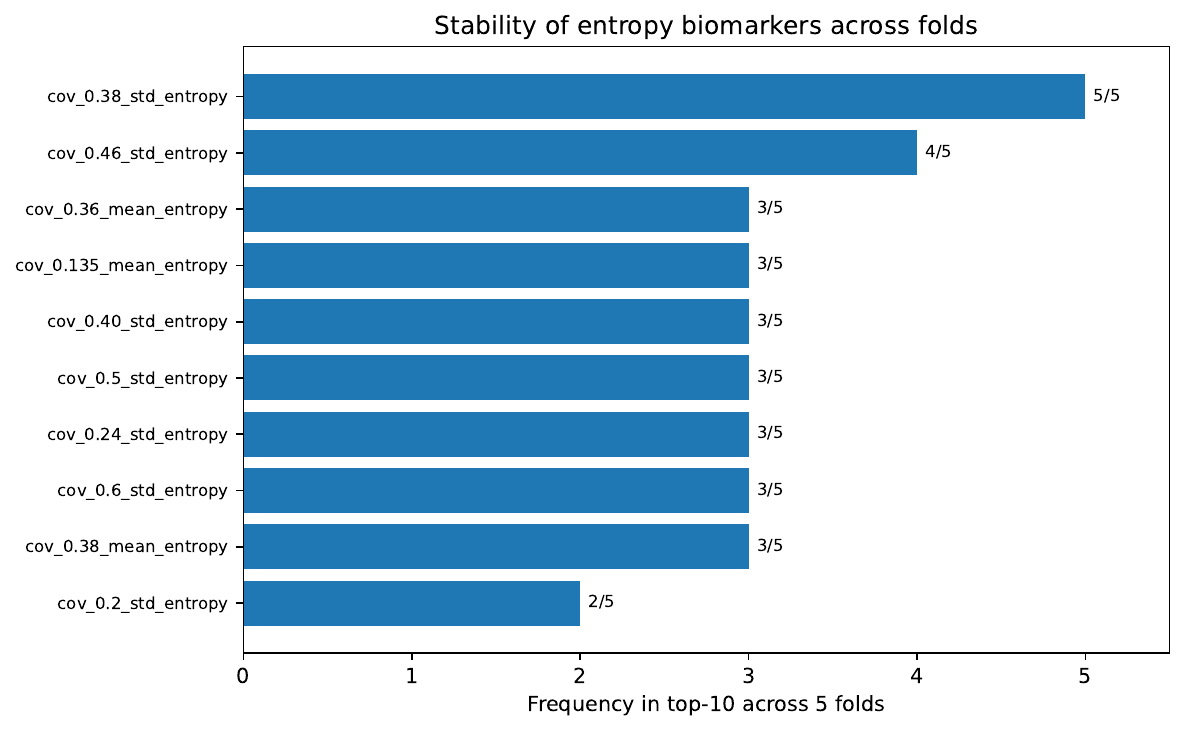}
\caption{Stability of top entropy biomarkers across folds.}
\end{figure}

\begin{figure}[H]
\centering
\includegraphics[width=0.85\linewidth]{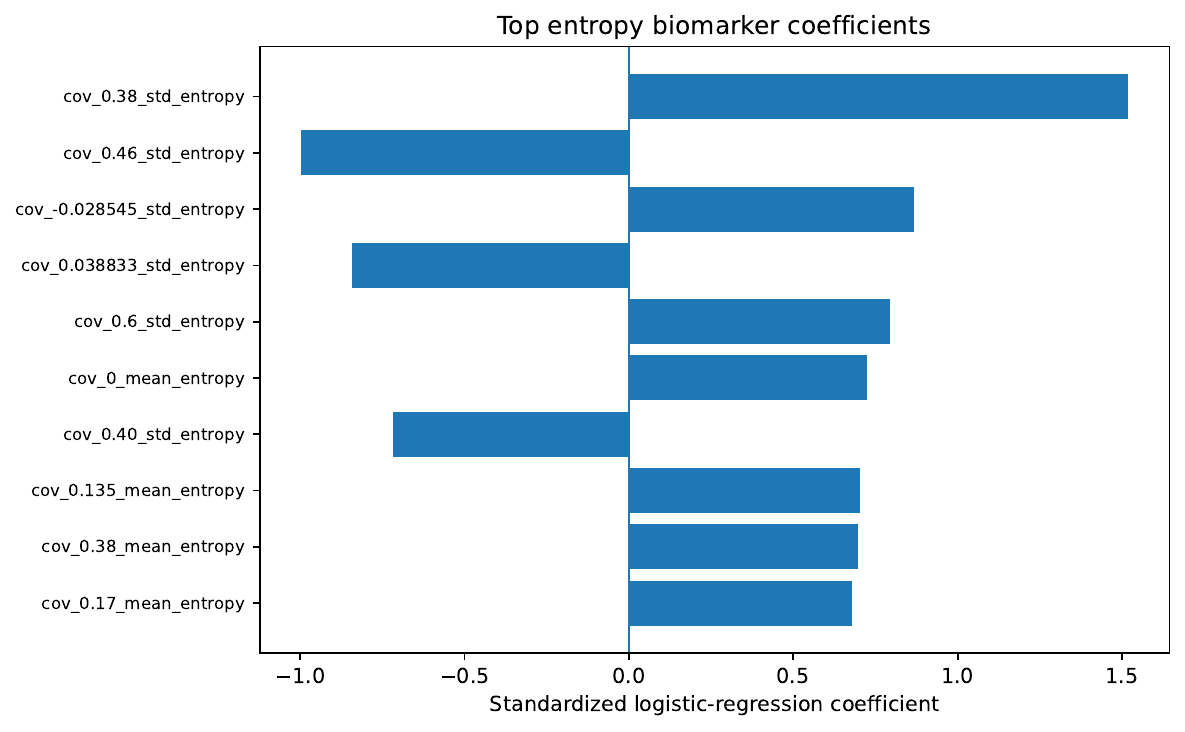}
\caption{Top standardized logistic-regression coefficients among entropy biomarkers.}
\end{figure}

\subsection{Negative results}

Sample entropy performed near chance (AUC 0.502), Higuchi fractal dimension performed below chance (AUC 0.464), and simple cross-modal coupling did not improve the entropy model. These findings suggest that the strongest signal in this dataset is not generic nonlinear complexity, but specifically distributional entropy of vocal trajectories.

\FloatBarrier

\section{Discussion}

This study found that entropy-driven vocal trajectory biomarkers outperform static pooling and several alternative complexity descriptors for depression detection from DAIC-WOZ interviews. The result was statistically significant under permutation testing and remained above the pooled baseline under nested cross-validation. The finding supports the hypothesis that depression-related information is expressed not only in average acoustic levels but also in temporal distributional uncertainty of vocal behavior.

Entropy biomarkers quantify how broadly and unpredictably a feature varies over the interview. The dominance of Shannon entropy over sample entropy and fractal dimension is informative. It suggests that discriminative signal is likely related to distributional variability across utterances rather than fine-grained deterministic pattern irregularity or scale-free fractal geometry. Several stable entropy biomarkers arose primarily from variability-related COVAREP channels, suggesting depression signal may be associated with dysregulated prosodic variability rather than mean acoustic level shifts.

RQA recurrence rate performed close to full trajectory dynamics but below entropy. Determinism performed poorly. This suggests recurrence density may capture useful state-repetition structure, whereas diagonal deterministic structure may be less stable in short utterance-level trajectories. Simple voice--gaze coupling showed modest standalone signal but did not improve entropy-based prediction.

The present model is best interpreted as a high-specificity risk stratification aid rather than a standalone screening tool. A high-specificity biomarker could help prioritize clinician review when combined with additional clinical data, but it should not be used as a standalone diagnostic tool.

\subsection{Translational deployment framework}

\begin{figure}[H]
\centering
\includegraphics[width=0.95\linewidth]{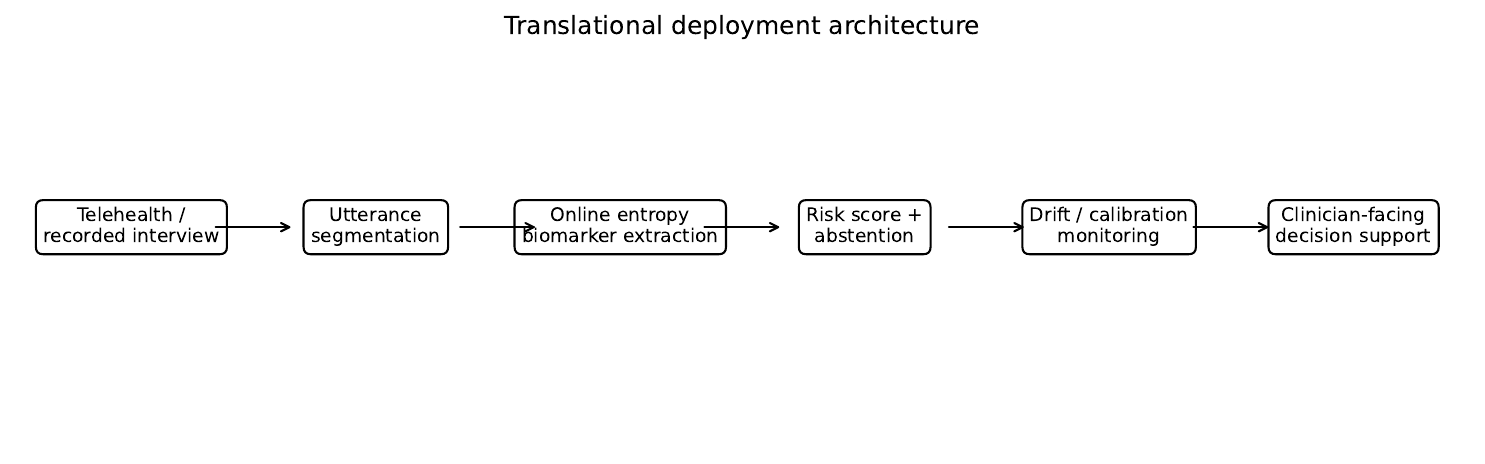}
\caption{Proposed translational deployment architecture. The model is intended as clinician-facing decision support rather than autonomous diagnosis.}
\label{fig:deploy}
\end{figure}

Figure~\ref{fig:deploy} illustrates a production-oriented architecture. In a deployed system, speech is collected during a telehealth or structured interview, segmented into participant utterances, transformed into online entropy biomarkers, and passed to a calibrated risk-scoring model. Safety controls should include confidence thresholds, abstention when data quality is poor, clinician-in-the-loop review, and monitoring for drift and calibration degradation.

\subsection{Methodological lessons}

A major lesson from this study is the importance of data-integrity checks in temporal biomarker modeling. The initial aggregate tensor contained duplicated participant trajectories, which would have produced misleading results. Rebuilding from raw participant files was essential. A second lesson is that post-hoc top-feature panels can overestimate performance. Leakage-aware nested evaluation and permutation testing are therefore necessary for credible small-sample biomarker discovery.

\subsection{Limitations}

This study has several limitations. First, the labeled cohort is modest in size, limiting statistical power and making external validation essential. Second, the current strongest model uses acoustic trajectory features; richer language features and self-supervised speech embeddings may provide complementary information. Third, stable COVAREP-derived markers should be mapped to official acoustic descriptors before final clinical interpretation. Fourth, the multimodal coupling analysis was preliminary and restricted to simple correlation-based features. Fifth, deep sequence models such as LSTM or attention networks were not fully benchmarked because the labeled sample size is small and overfitting risk is high.

\subsection{Future work}

Future studies should validate entropy biomarkers on independent cohorts, map stable acoustic entropy markers to interpretable prosodic descriptors, evaluate transfer entropy and nonlinear multimodal coupling, and test lightweight attention models under strict nested validation. Longitudinal studies could determine whether entropy biomarkers track symptom change over time.

\section{Reproducibility}

The analysis was implemented in Python using scikit-learn and standard scientific-computing libraries \cite{pedregosa2011scikit}. The pipeline includes raw-file reconstruction, utterance-level alignment, biomarker extraction, cross-validation, nested validation, permutation testing, and figure generation. Random seeds were fixed where applicable. Code and derived non-identifying outputs should be released with the final manuscript subject to DAIC-WOZ data-use restrictions.

\section{Ethics Statement}

This study uses an existing research corpus distributed for scientific use. The proposed model is intended for research and decision support, not autonomous diagnosis. Any deployment requires prospective validation, privacy protection, demographic bias assessment, clinician oversight, and appropriate regulatory review.

\section{Conclusion}

Entropy-dominated temporal vocal biomarkers provide statistically significant and interpretable signal for depression detection beyond static temporal aggregation. These results suggest depression-related signal may lie less in mean acoustic behavior than in entropy of behavioral dynamics. Compared with recurrence, sample entropy, fractal dimension, and simple multimodal coupling, Shannon entropy of vocal trajectories emerged as the dominant biomarker family. These findings support temporal distributional uncertainty in conversational speech as a promising digital phenotype for depression-related assessment.

\section*{Declaration of Competing Interest}
The author declares no competing interests.

\section*{Data and Code Availability}
The DAIC-WOZ dataset is available under its original access conditions. Code for feature reconstruction, biomarker extraction, validation, and figure generation is intended for public release in a repository accompanying the manuscript, excluding protected raw data.

\section*{Author Contributions}
H.S.S. conceived the study, implemented the analysis pipeline, performed experiments, interpreted results, and wrote the manuscript.

\section*{Acknowledgments}
The author acknowledges the creators and maintainers of the DAIC-WOZ dataset and the developers of open-source scientific Python software.

\bibliographystyle{elsarticle-num}
\bibliography{references}

\end{document}